\documentclass[aps,twocolumn, showpacs]{revtex4}
\usepackage{graphicx}
\usepackage{amsfonts} 
\usepackage{amssymb,amsmath}

\begin{document}
\title{Distance growth of quantum states due to  initial system--environment correlations}
 \author{J. Dajka and J.  \L uczka} 
\address{Institute of Physics, University of Silesia, 40-007 Katowice, Poland }

\begin{abstract} 
Intriguing  features of the distance between two arbitrary states  of  an open quantum system are identified that are  induced by initial  system-environment correlations. As an example, we analyze a qubit dephasingly coupled to a bosonic environment.  Within tailored parameter regimes, initial correlations are shown to substantially increase a distance between two qubit states evolving to long-time limit  states according to  exact non-Markovian dynamics.    It exemplifies the breakdown of the distance contractivity of  the reduced dynamics. 
\end{abstract}


\maketitle
%
{\it Introduction}. --There are various states of physical systems changing  and evolving in time.  Thermodynamic states or  quantum states are well known examples. One can pose natural questions: How "similar" is a given nonequilibrium state to the equilibrium one?  How "similar" is an entangled quantum state to the unentangled one? Is there a natural quantifier to measure the separation  between  states?  The answer is not unique and depends on a character of a problem which we  analyze.  In the simplest way, we could say that two states are close to each other if averaged values of observables in these states are close to each other.  For quantum systems,  there  is a natural distance induced by a norm which in turn is induced by the scalar product of vectors in Hilbert space of a given quantum system.  One of the central issues of  quantum engineering is to prepare a quantum system  in a  state having  desired features. Doing this one is faced with many difficulties: unavoidable decoherence, imperfect implementation of quantum operations and many other sources of unpredictable mess. 
The statistical similarity  of   quantum states  can be  related to certain measures of a distance between them \cite{gil}. One of the most popular quantifiers for the  distiguishability of the probability distributions arising from the measurements conducted on  quantum states (or shortly 'distinguishability of states') is the trace distance \cite{nielsen}. The trace distance  of any two states (density matrices) $\rho_1$ and $\rho_2$  is defined as  \cite{nielsen} 
\begin{eqnarray}\label{tracegen}
D[\rho_1, \rho_2]=|| \rho_1-\rho_2||,  
\end{eqnarray}
where for any operator $A$ its norm is defined by the relation
%
$|| A|| =(1/2) \, \mbox{Tr} \sqrt{  A^{\dagger} A }$.
%
Any positive and trace-preserving map $\cal E$ defined on the whole space of  trace class operators on the Hilbert space   is contractive, i.e. 
\begin{eqnarray}\label{contr}
D[{\cal E}(\rho_1),  {\cal E}(\rho_2) ] \le D[\rho_1, \rho_2].
\end{eqnarray}
In particular, when ${\cal E} = {\cal E}_t$ is a completely positive quantum dynamical semigroup  such that $\rho(t) = {\cal E}_t \rho(0)$, then 
\begin{eqnarray}\label{contr}
D[\rho_1(t) , \rho_2(t)  ] \le D[\rho_1(s), \rho_2(s)]  \quad \mbox{for} \quad t > s, 
\end{eqnarray}
i.e. the distance cannot increase in time.  It means that 
the  distinguishability of any states  can not increase above an initial value.    Motivated by   
findings recently reported in Ref. \cite{breu3},  
 we study  an open quantum  system  which is initially correlated with its environment 
\cite{pech}. 
 We consider a  dephasing model \cite{defaz,devoret} of decoherence of a qubit interacting with a  bosonic environment.  If the qubit and its environment is initially prepared in an uncorrelated  state, the reduced dynamics is completely positive and hence contractive. In the presence of initial correlations one is left with positive (but non completely positive) dynamics \cite{zycz,ban} and the contractivity may fail \cite{breu3}. The   model of dephasing  allows to find an exact reduced dynamics and reveals highly non-trivial properties of distinguishability with respect to the trace distance.  
 We explicitly show that under tailored regimes the contractivity fails for some initially entangled states. The trace distance of different states  grows above its initial value even in the long time limit.  In consequence, the distinguishability of the long-time limit states can increase above its initial value.  

 {\it Model}. --The system we study is  a qubit $Q$, formed by an arbitrary two-level system  coupled to its  environment $B$.  
 We consider the case when the process of energy dissipation  is negligible and only pure decoherence  is a mechanism for decoherence of the qubit.   
It leads to an irreversible process of information  loss.  
We model  such a system  by the Hamiltonian  ($\hbar=1$)
\begin{eqnarray}\label{ham} 
H= H_Q \otimes  {\mathbb{I}}_B +{\mathbb{I}}_Q  \otimes H_B + S^z \otimes H_I, \\
H_Q = \varepsilon S^z, \quad  
H_B=\int_0^\infty d\omega \, h( \omega) a^\dagger(\omega)a(\omega), \\
H_I=\int_0^\infty d\omega  \left[ g^*(\omega)a(\omega) +g(\omega)a^\dagger(\omega)\right],  
\end{eqnarray}
where  $S^z$  is the z-component of the spin operator and is represented by the diagonal matrix $S^z =diag[1,-1]$ of elements $1$ and $-1$. The parameter  $ \varepsilon$ is the  qubit energy splitting,   ${\mathbb{I}}_Q$   and ${ \mathbb{I}}_B$  are   identity operators (matrices) in corresponding Hilbert spaces of the  qubit $Q$ and the environment $B$, respectively.  
The operators $ a^\dagger(\omega)$ and $a(\omega)$ are the bosonic creation and annihilation operators, respectively.  The real-valued spectrum function $h(\omega)$  characterizes the environment. The coupling is described  by the function $g(\omega)$ and  the function 
 $ g^*(\omega)$ is   the complex conjugate to   $g(\omega)$. 
The Hamiltonian (\ref{ham}) can be rewritten in the  block--diagonal structure \cite{fidel}, 
\begin{eqnarray}
H=diag[H_{+}, H_{-}],  \quad 
H_\pm =  H_B  \pm   H_I  \pm  \varepsilon { \mathbb{I}}_B. 
\end{eqnarray}
 We consider  a correlated  initial state of the qubit--environment  system in the form  similar to that 
in  Ref.  \cite{kossak}, namely,  
\begin{eqnarray}\label{ini}
|\Psi(0)\rangle =b_+|1\rangle \otimes |\Omega_0\rangle +b_-|-1\rangle \otimes 
|\Omega_{\lambda}\rangle.  
\end{eqnarray}
The states $|1\rangle$ and $|-1\rangle$ denote the excited and
ground states of the qubit, respectively, the non-zero complex numbers $b_+$ and $b_-$ are such that 
$|b_+|^2 + |b_-|^2 =1$.  The states    
 $|\Omega_0\rangle$   and $|\Omega_{\lambda}\rangle$ are  environment states.  As an example, let us assume that  $|\Omega_0\rangle$
is an environment ground state  and   
\begin{eqnarray}\label{omega1}
|\Omega_{\lambda} \rangle = C_{\lambda}^{-1} \, \left[(1-\lambda)|\Omega_0\rangle +\lambda |\Omega_f\rangle  \right] 
\end{eqnarray}
is a linear combination of a ground state $|\Omega_0\rangle$ and the coherent state 
%
$|\Omega_f\rangle =D(f) |\Omega_0\rangle$,  
%
where the displacement operator $D(f) $  reads \cite{brat} 
\begin{eqnarray}\label{displacement}
D(f)=\exp\left\{\int_0^\infty d\omega \left[ f(\omega)a^{\dagger}(\omega) -  f^*(\omega)a(\omega)\right]\right\} 
\end{eqnarray}
 for an arbitrary square--integrable function $f$.  
The constant  $C_{\lambda}$ normalizes the state $|\Omega_{\lambda} \rangle$ and reads 
\begin{eqnarray}\label{C}
C_{\lambda}^2=(1-\lambda)^2+\lambda^2+2\lambda(1-\lambda) 
Re \langle \Omega_0|\Omega_f\rangle,    
\end{eqnarray}
where ${Re}$ is a real part of the scalar product $\langle \Omega_0|\Omega_f\rangle $ of two states in the environment Hilbert space.  
The parameter $\lambda\in[0,1]$  controls   the initial entanglement of the qubit and  environment.  For $\lambda=0$ the qubit  and environment are initially uncorrelated while for  $\lambda=1$ the entanglement is, for a given class of initial states, maximal. The states with $\lambda=1$ are not maximally entangled in the usual sense as the coherent states, forming an overcomplete set of states, are not mutually orthogonal. 
 The state of the total system  at time $t>0$  has the form      
\begin{eqnarray}  \label{ewol}
|\Psi(t)\rangle =b_+|1\rangle \otimes |\psi_+(t)\rangle +b_-|-1\rangle \otimes 
|\psi_-(t)\rangle,  
\end{eqnarray}
where $|\psi_+(t)\rangle=\exp(-i H_+ t)   |\Omega_0\rangle$ and $|\psi_-(t)\rangle=\exp(-i H_- t)   |\Omega_{\lambda}\rangle$. 
The density  matrix $\rho_{\lambda}(t)$ of the qubit is obtained from the relation 
\begin{eqnarray}  \label{rho}
\rho_{\lambda}(t) =  \mbox{Tr}_B \left\{|\Psi(t) \rangle \langle\Psi(t)| \right\}
\end{eqnarray}
and  $\mbox{Tr}_B$ denotes the partial tracing over the environment. Its explicit form reads 
\begin{eqnarray}\label{ro}
\rho_{\lambda}(t)=\left(\begin{array}{cc} |b_+|^2 & b_+b_-^*A_{\lambda}(t) \\ b_+^*b_-A_{\lambda}^*(t) & |b_-|^2 \end{array}  \right), 
\end{eqnarray}
%
%
\begin{eqnarray}\label{A}
 A_{\lambda}(t)= C_{\lambda}^{-1} \,e^{-2i\varepsilon t} e^{- r(t)} \, \left[1-\lambda+\lambda 
e^{-2i\Phi(t)} e^{ s(t)}  \right],  
\end{eqnarray}
where \cite{fidel} 
\begin{eqnarray}\label{U}
r(t) = 4\int_0^\infty d\omega g_h^2(\omega) \left[1-\cos(\omega t) \right], \quad \quad\quad \quad\quad\nonumber\\
s(t)=2\int_0^\infty d\omega g_h(\omega)f(\omega)\left[1-\cos(\omega t)\right]  
- \frac{1}{2} \int_0^\infty d\omega f^2(\omega) ,  \quad \nonumber
\end{eqnarray}
%
where $g_h(\omega) = g(\omega)/h(\omega)$ and   
\begin{eqnarray}\label{Fi}
\Phi(t) =  \int_0^\infty d\omega g_h(\omega)f(\omega) \sin(\omega t).
\end{eqnarray}
For the sake of simplicity, we have assumed that the  functions $g(\omega)$ and $f(\omega)$ are real. 
The  modeling of the qubit coupling to the environment is performed  in terms of the spectral density function 
\begin{eqnarray}\label{J}
g_h^2(\omega)=\alpha  \, \omega^{\mu-1}\exp(-\omega/\omega_c), 
\end{eqnarray}
where $\alpha >0$ is the qubit-environment coupling constant, $\mu>-1$ is the "ohmicity" parameter 
(the case $\mu =0$  corresponds to the ohmic  and $\mu >0$  to super-ohmic environments, respectively) and 
$\omega_c$ is a cut-off  frequency. 
The coherent state $|\Omega_f\rangle$  is determined by the function 
\begin{eqnarray}\label{f}
f^2(\omega)&=&\gamma \, \omega^{\nu-1}\exp(-\omega/\omega_c). 
\end{eqnarray}
This choice is arbitrary but convenient as it allows for finding an explicte analytic formulas   for 
\begin{eqnarray}\label{LL}
r(t)=4\mathcal{L}(\alpha,\mu,t)],   \quad \quad  \quad     \nonumber \\
s(t)= 2\mathcal{L}(\sqrt{\alpha \gamma},(\mu+\nu)/2,t)-\gamma\Gamma(\nu)\omega_c^\nu, 
  \nonumber \\
\mathcal{L}(\alpha,\mu,t) =\lambda\Gamma(\nu) \omega_c^\mu\left\{1-\frac{\cos\left[\mu\arctan(\omega_c t) \right] }{(1+\omega_c^2t^2)^{\mu/2}}\right\}, \nonumber\\
\Phi(t) = \sqrt{\alpha \gamma}\; \Gamma(\kappa) \omega_c^{\kappa} 
\;  \frac{\sin\left[\kappa \arctan(\omega_c t) \right] }{(1+\omega_c^2t^2)^{\kappa/2}}, \quad
\end{eqnarray}
where $\Gamma(\nu)$ is the Euler gamma function and the parameter $\kappa =(\mu+\nu)/2$. 

{\it Distance of states}. --Let us  examine  the distinguishability between  two various, initially correlated ($\lambda\neq 0$) states.  The corresponding trace distance reads   
\begin{eqnarray}\label{D2}
 D^2[\rho_{\lambda_{1}}(t), \rho_{\lambda_{2}}(t)]  = \left(|b_+^{(1)}|^2 - |b_+^{(2)}|^2\right)^2
\nonumber \\
+ |b_+^{(1)} b_-^{(1)*} A_{\lambda_{1}}(t) - b_+^{(2)} b_-^{(2)*} A_{\lambda_{2}}(t)|^2. 
\end{eqnarray}
As the initial  state  (\ref{ini}) of the qubit--environment composite system is generically entangled,   the non--local operations are involved in its preparation.  Such a procedure would essentially   require highly sophisticated quantum  engineering. Naively speaking one can recognize the following steps: {\it (i)} the first, when one prepares the state $|\Omega_\lambda\rangle$ given by Eq. (\ref{omega1}), next  "tensorize" it with the ground state of the qubit $|-1\rangle$; {\it (ii)}  the second step, when one superpose, with the weights $b_\pm$, the result of the first step with the excited qubit $|1\rangle$ state tensorized with the vacuum $|\Omega_0\rangle$. Both steps demand using precise technics and their details are essentially beyond the scope of this paper. A different preparations of  initial states (\ref{ini}) can be provided both via changing details of the first step and the second step of the procedure.   
In other words, one can change the parameters $b_\pm$ in (\ref{ini}) or  manipulate 
on the state $|\Omega_\lambda\rangle$ (by changing $\lambda$).
When two different states are determined by two different sets of numbers $b_\pm^{(k)} \;(k=1, 2)$ in the superposition  (\ref{ini})  with the same state $|\Omega_\lambda\rangle$
then $ A_{\lambda_{1}}(t)  =  A_{\lambda_{2}}(t)$ and Eq. (\ref{D2}) takes the form   
\begin{eqnarray}\label{D_1}
 D^2[\rho_{\lambda_{1}}(t), \rho_{\lambda_{2}}(t)]  = \left(|b_+^{(1)}|^2 - |b_+^{(2)}|^2\right)^2
\nonumber \\
+|b_+^{(1)} b_-^{(1)*} - b_+^{(2)} b_-^{(2)*}|^2 \;  |A_{\lambda_{1}}(t)|^2. 
\end{eqnarray}
The function $| A_{\lambda_{1}}(t)|$ is a decaying  function of time and   
the distance  (\ref{D_1}) also  decays as time grows. In consequence  the distinguishability between  two  initially correlated states always becomes lowered.  It is clear that no matter what $b_\pm^{(k)}$  are, the reduced dynamics results in contraction of the distance between the states.
So, the only chance  to observe  the growth of the distinguishability of qubit reduced states is  to modify the parameters of environment encoded in $|\Omega_\lambda\rangle$ in Eq.(\ref{omega1}). 
\begin{figure}[htpb]
\includegraphics[width=0.55\textwidth,angle=0]{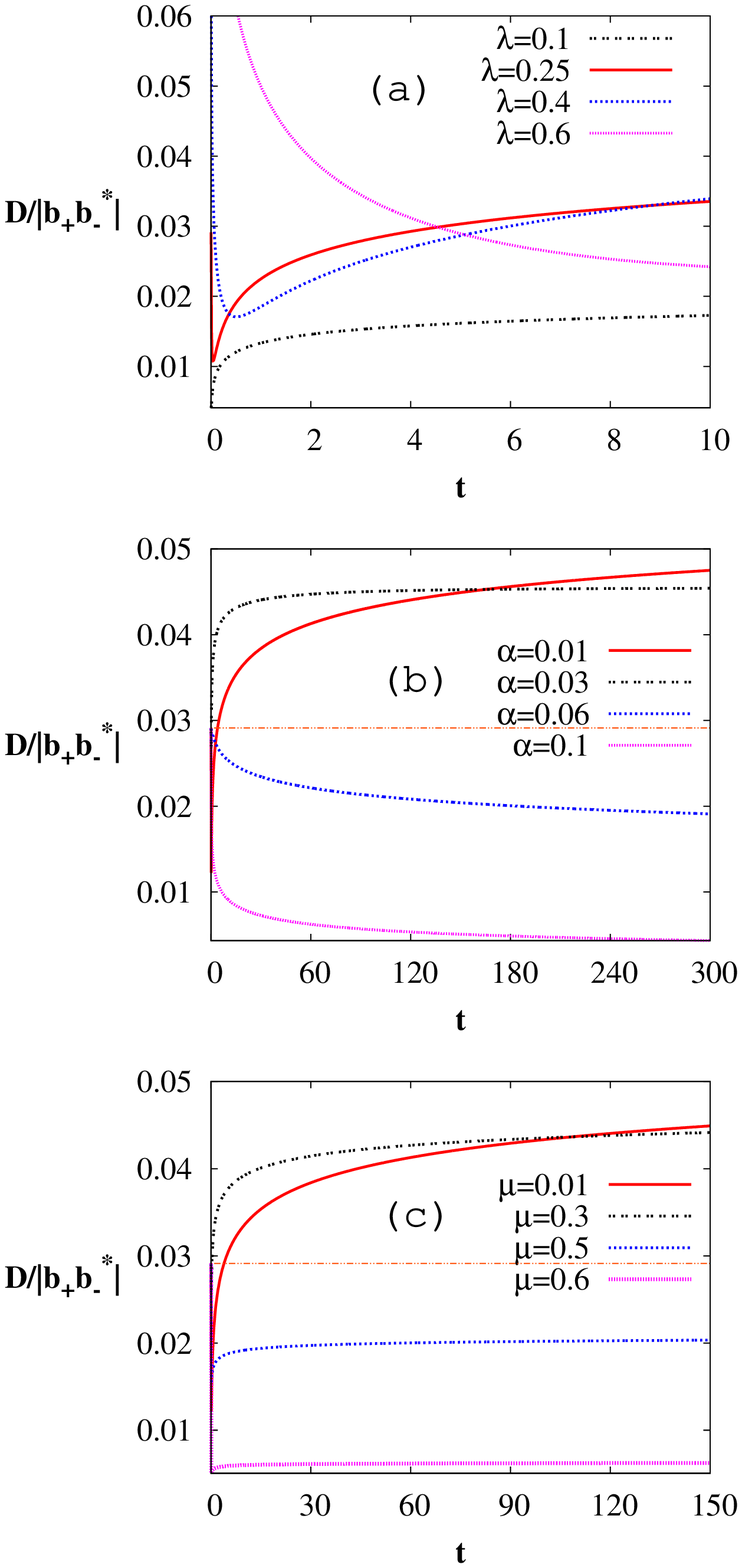}
\caption{Time evolution of the trace distance $D[\rho_{\lambda}(t), \rho_{0}(t)]$ between the initially correlated and non-correlated states for selected values of parameters.  Dimensionless time is in unit of $\omega_c$.  The remaining parameters are: 
  $\varepsilon =1, 
(a) \mu=0.01$, $\nu=0.05$, $\alpha = 0.01, \gamma=0.05$;  
(b) $ \mu=0.01$, $\nu=0.05$ ,  $ \gamma=0.05$. $\lambda = 0.25$;  
(c)   $\nu=0.05$ ,  $ \gamma=0.05$. $\lambda = 0.25$,       $\alpha =0.01$. 
The horizontal lines in (b) and (c) indicate the initial distance. 
}
\label{fig1}
\end{figure}
Let us consider this case assuming that  
$b_\pm^{(k)}= b_\pm$ are fixed and  the same  for  $k=1, 2$,  and 
 two different states are determined by two different sets of numbers $\lambda_1$ and $\lambda_2$. 
In such a case the distance reads 
\begin{eqnarray}\label{2lamda}
 |b_+b_{-}^{*}|^{-1} \; D[\rho_{\lambda_{1}}(t), \rho_{\lambda_{2}}(t)]  = 
|A_{\lambda_{1}}(t)-A_{\lambda_{2}}(t)| 
\nonumber \\
= e^{-r(t)} \left\{a^2  + b^2 e^{2 s(t)} + 2 a b e^{ s(t)} \cos[2\Phi(t)]\right\}^{1/2}, 
\end{eqnarray}
where 
\begin{eqnarray}\label{ab}
 a=\frac{1-\lambda_1}{C_{\lambda_1}} - \frac{1-\lambda_2}{C_{\lambda_2}}  , \quad 
b=\frac{\lambda_1}{C_{\lambda_1}} - \frac{\lambda_2}{C_{\lambda_2}}. 
\end{eqnarray}
In Fig. \ref{fig1}(a), we show time evolution of the trace distance $D[\rho_{\lambda}(t), \rho_{0}(t)]$ between the initially correlated and non-correlated states for four selected degrees of correlation determined by the value of the correlation parameter $\lambda$.  At time $t=0$, the distance 
$D=0.004, 0.03, 0.08, 0.19 $ for $\lambda =0.1, 0.25, 0.4, 0.6$, respectively.  In the long time limit,  the distance between states takes the values $D= 0.03, 0.08, 0.12, 0.15$,  respectively.  In the first three cases, the distance between long-time limit states  is greater than the initial distance and the distinguishability of final states is better than initial states.    There is some optimal correlation $\lambda$  for which the   distinguishability  of  final  states is the best.  
However, if the initial correlation is strong enough, the distance for  long time  is smaller than at the initial time.  It is important observation that there exists a critical value  $\lambda_c$ (depending on other parameters) such that for $\lambda > \lambda_c$ the stationary states are less distinguishable than initial states.  In some regimes (as e.g. the case  $\lambda = 0.25, 0.4$ in Fig. \ref{fig1}(a)), at  early stage of evolution, the distance decreases reaching a minimum and next it  increases and saturates  for long time. 
In  Fig. \ref{fig1}(b) and Fig. \ref{fig1}(c) we  show how the distance can depend on environment characteristics: the coupling constant  $\alpha$ and the "super-ohmicity" $ \mu >0$.   It follows that strong coupling of the qubit to environment diminishes the distinguishability  of initial states. Similarly, 
if the environment is  more super-ohmic, the distinguishability of states is weaker.   
\begin{figure}[t]
\begin{center}
\includegraphics[width=0.3\textwidth, angle=270]{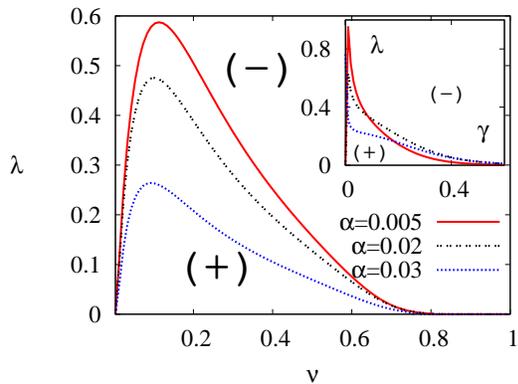}
\end{center}
\caption{In the region  limited by the curves and the  horizontal axes   (denoted by $+$),  
$\lim_{t\to\infty} D[\rho_{\lambda_1}(t), \rho_{0}(t)] > D[\rho_{\lambda_1}(0), \rho_{0}(0)]$.  Outside this region (denoted by $-$),   the distance between two states at time $t=0$ is greater than between  the long-time limit states (the contractive region).  The parameters are: $\varepsilon =1$, 
 $\gamma=0.05, \mu =0.01$; 
Inset: $\mu=0.01, \nu=0.05$. 
}
\label{fig2}
\end{figure}
%
%
In Fig. \ref{fig2},  the influence of  parameters characterizing initial environment state is depicted.  
Both regions, where the  "distinguishability gain" (i. e.  the distance in the long time limit is greater than for the initial time) is reached, are limited 
and located in the corner of the corresponding parameter planes. It means that not all  but specific environments  can induce the "distinguishability gain".

Finally,  we consider the case when both $\lambda_1$ and $\lambda_2$ are non-zero, see Fig. 3.  At time $t=0$, the distance  is 
$D=0.015, 0.014, 0.0071, 0.026 $ for $\lambda =0 , 0.05, 0.3, 0.4$, respectively.  In the long time limit,  the distance takes the values $D= 0.039, 0.031, 0.0072, 0.020$,  respectively. In the first three  cases, the final distance is greater than the initial distance and the distinguishability of final states is better than initial states.   
The region of  the $(\lambda_1, \lambda_2)$-plane, where this effect occurs is bounded by 
$\lambda_1$ and $ \lambda_2$ axes and a decreasing function $\lambda_2$ {\it vs} $ \lambda_1$. 
 From the presented  example   we detect that if  $\lambda_2$ exceeds some fixed value  of  $\lambda_1$, this effect is distinctly absent  and the distance  in the long time limit  can  never be  greater than for $t=0$. \\
\begin{figure}[tb]
\begin{center}
\includegraphics[width=0.30\textwidth,angle=270]{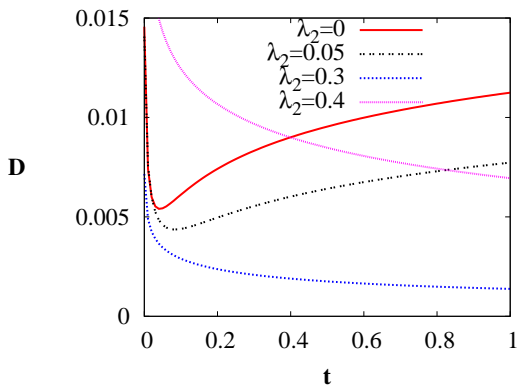}
\end{center}
\caption{Time evolution of the trace distance $D[\rho_{\lambda_1}(t), \rho_{\lambda_2}(t)]$  for 
fixed $\lambda_1 = 0.25$ and several values of $\lambda_2$.  Time is scaled as in Fig. 1. 
 The parameters  $\mu=0.01$, $\nu=0.05$ $\alpha=0.01$,   $\gamma = 0.25$  and $b_+=b_-=1/\sqrt{2}$. }
\label{fig3}
\end{figure}
{\it Summary}. --Recent investigations  have often shown that   dynamics of  quantum systems can exhibit various counter--intuitive and non--trivial features which can be inferred from detailed analysis of highly sophisticated theoretical models. It is of great importance to  verify the predictions in a carefully prepared experiments. An effective design of such experiments would require guidelines provided by theoretical studies of realistic physical models. Our work reported in this paper, as a step in this direction,  could be placed somewhere in between: we consider a fairly simple model of decoherence under very specific conditions but present results which are exact. Despite its simplicity the model is realistic enough to be  experimentally accessible \cite{devoret}. According to common wisdom the decoherence resulting from the "openness" of quantum systems causes blurring of the information encoded in quantum states: as it has recently   been shown \cite{breu3} in the context of distinguishability of quantum states, it is not always the case.  In this paper we provide an explicit realization of the recent suggestion on the distinguishability growth due to initial qubit--environment entanglement. The distinguishability growth occurs not only at the short time scales but is shown  to be a feature of  long-time limit states. 
 This feature seems to be favorable for the potential experimental verification of the predictions presented in \cite{breu3} especially when the desired short--time growth would occur in ultra--short time scales. We have shown that the induced distinguishability growth is not generic for the considered model and  we identified the parameter  regimes  where the effect is the most apparent.

The work  supported by   the  ESF Program  "Exploring the Physics of Small Devices".  
 

\begin{thebibliography}{50}
\bibitem{gil} A. Gilchrist {\it et al.}, Phys. Rev. A {\bf 71}, 062310 (2005). 
\bibitem{nielsen} M. A. Nielsen and L. I.  Chuang, {\it Quantum  Computation and Quantum Information} 
(Cambridge University Press, Cambridge, U.K., 2000). 

\bibitem{breu3} E.--M. Laine {\it et al.}, arxiv:1004.2184v1 (2010). 

\bibitem{pech} P. Pechukas, Phys. Rev. Lett. 7{\bf 3}, 1060 (1994); 
P. Stelmachovic and V. Buzek, Phys. Rev. A {\bf 64}, 062106 (2001); 
N. Boulant {\it et al.},  J. Chem. Phys. {\bf 121}, 2955 (2004); 
 F. Jordan {\it et al.}, Phys. Rev. A {\bf 70}, 052110 (2004).



\bibitem{devoret} {D. I. Schuster  {\it et al.}},  {Nature} {\bf 445} 515 (2007).

\bibitem{defaz}  J. {\L}uczka, Physica A  {\bf 167}, 919 (1990). 


\bibitem{ban} M. Ban, Phys. Rev.  A {\bf 80}, 064103 (2009).

\bibitem{zycz} H. A. Carteret {\it et al.}, Phys. Rev. A {\bf 77}, 042113 (2008).

\bibitem{fidel}  J. Dajka and J. {\L}uczka, Phys. Rev. A {\bf 77}, 062303 (2008); 
 J. Dajka {\it et al.}, Phys. Rev. A {\bf 79}, 012104 (2009). 


\bibitem{kossak} P. Anielo  {\it et al.},  Open Sys. \& Inf. Dyn. (accepted),  arXiv:0912.4123 (2009).

\bibitem{brat} O. Brattelli and  D. W.  Robinson, {\it Operator Algebras and Quantum Statistical Mechanics 2} (Springer, Berlin, 1997).


\end{thebibliography}
\end{document}